# The decay mode of high-energy tails in velocity distributions of astrophysical plasma particles

Jian-Miin Liu*

Department of Physics, Nanjing University

Nanjing, The People's Republic of China

*On leave. E-mail address: liu@phys.uri.edu.

ABSTRACT

The relativistic equilibrium velocity distribution coincides with the Maxwellian distribution for small velocities and vanishes at c, the velocity of light. Based on the decay pattern of high-energy tail in the relativistic equilibrium velocity distribution, it is predicted for velocity and velocity rate distributions of astrophysical plasma particles that they fall off to zero, as velocity goes to c, slower than any exponential decay but faster than any power-law decay.

In statistical mechanics based on pre-relativistic mechanics, velocity and velocity rate distributions of particles in equilibrium are known to be the Maxwellian,

$$M(y^1,y^2,y^3)dy^1dy^2dy^3 = N(\frac{m}{2\pi K_B T})^{3/2} \exp[-\frac{m}{2K_B T}(y^2)]dy^1dy^2dy^3, \qquad (1a)$$

$$M(y)dy = 4\pi N(\frac{m}{2\pi K_B T})^{3/2}(y^2)\exp[-\frac{m}{2K_B T}(y^2)]dy, \qquad (1b)$$

where N is the number of particles, m their rest mass, T the temperature, and $K_B$ the Boltzmann constant, $y=(y^r y^r)^{1/2}$, r=1,2,3, $y^r$ is the well-defined Newtonian velocity. Eqs.(1a-1b) indicate the Gaussian decay of high-energy tail in velocity and velocity rate distributions.

However, it has been observed for many years that velocity distributions of astrophysical plasma (planetary magnetospheres, solar wind and other) particles deviate from the Maxwellian in the high-velocity part though coincide with the Maxwellian in the low-velocity part [1-3]. Experimental data clearly showed some slower decay of high-energy tails than the Gaussian or any other exponential decay. As experimental data seem to be well modeled by the κ (kappa) distribution, as the kappa distribution implies a power-law decay for a fixed value of kappa, some authors concluded that the decay mode of high-energy tails in velocity distributions is power-law like. This conclusion is rather misleading.

The kappa distribution has the shape of [3]

$$K(y)dy = \frac{N}{\pi^{3/2}} \frac{1}{\theta^3} \frac{\Gamma(\kappa+1)}{\kappa^{3/2}\Gamma(\kappa-1/2)} (1+\frac{y^2}{\kappa\theta^2})^{-(\kappa+1)}dy \qquad (2)$$

where $\theta = [(2\kappa-3)/\kappa]^{1/2}(K_BT/m)^{1/2}$, $\Gamma$ is the gamma function and kappa is a parameter to be determined in comparison with experimental data. Different values of kappa correspond to different kinds of velocity distribution. When and only when kappa goes to infinity, the kappa distribution becomes the Maxwellian. For any fixed value of kappa, the kappa distribution extends as far as y=$\infty$, while experimental data, we believe, extend as far as y=c, where c is the velocity of light. Also, for any fixed value of kappa, the kappa distribution does not reduce to the Maxwellian in the low-velocity part, while experimental data, as observed, can be well described by the Maxwellian distribution in the low-velocity part. The kappa distribution can not be a good modeling distribution for experimental data. Moreover, the kappa distribution is phenomenological: the value of kappa in fitting experimental data varies from event to event [2].

In our recent work [4], we made the relativistic corrections to Maxwell's velocity distribution law and obtained the relativistic equilibrium velocity and velocity rate distributions. They provide with the appropriate modeling distributions.

The relativistic equilibrium velocity and velocity rate distributions are

$$P(y^1,y^2,y^3)dy^1dy^2dy^3 = N\frac{(m/2\pi K_BT)^{3/2}}{(1-y^2/c^2)^2} \exp[-\frac{mc^2}{8K_BT}(\ln\frac{c+y}{c-y})^2]dy^1dy^2dy^3, \qquad (3a)$$

$$P(y)dy = \pi c^2 N\frac{(m/2\pi K_BT)^{3/2}}{(1-y^2/c^2)} (\ln\frac{c+y}{c-y})^2 \exp[-\frac{mc^2}{8K_BT}(\ln\frac{c+y}{c-y})^2]dy. \qquad (3b)$$

Eqs.(3a-3b) are so-named because they are based on the velocity space in special relativity and the modified special relativity theory [5].

$P(y^1,y^2,y^3)$ and $P(y)$ respectively reduce to the Maxwellian velocity and velocity rate distributions for small velocities, y/c <<1; They both, as y goes to c, fall off to zero slower than any exponential decay but faster than any power law decay.

Actually, $M(y^1,y^2,y^3)$ and $M(y)$ are respectively the first-order approximations of $P(y^1,y^2,y^3)$ and $P(y)$. That proves the first part of the statement. To judge its second part, we first consider three decay modes, $\exp\{-A[\ln(2cz)]^2\}$, $z^{-n}$, and $\exp\{-[2cz]^B\}$, where z is a variable going to infinity, A is a positive constant, B and n are two positive numbers. One can find

$$P = \lim_{z\to+\infty} \frac{\exp\{-A[\ell n(2cz)]^2\}}{z^{-n}} = P\lim_{z\to+\infty} \frac{n}{2A\ell n(2cz)}, \qquad (4a)$$

which gives

$$P=0. \qquad (4b)$$

One can also find

$$Q = \lim_{z \to +\infty} \frac{\exp\{-[2cz]^B\}}{\exp\{-A[\ell n(2cz)]^2\}} = Q \lim_{z \to +\infty} \frac{2A\ell n(2cz)}{B(2cz)^B}, \qquad (5a)$$

and

$$Q = 0. \qquad (5b)$$

We now rewrite $\lim\limits_{y \to c} P(y^1,y^2,y^3)$ and $\lim\limits_{y \to c} P(y)$ as

$$\lim_{y \to c} P(y^1,y^2,y^3) = \lim_{z \to +\infty} \frac{c^2 N}{4} \left(\frac{m}{2\pi K_B T}\right)^{3/2} z^2 \exp\{-A[\ell n(2cz)]^2\}, \qquad (6a)$$

$$\lim_{y \to c} P(y) = \lim_{z \to +\infty} \frac{\pi c^3 N}{2} \left(\frac{m}{2\pi K_B T}\right)^{3/2} z [\ell n(2cz)]^2 \exp\{-A[\ell n(2cz)]^2\}, \qquad (6b)$$

where $z = 1/(c-y)$ and $A = \dfrac{mc^2}{8K_B T}$. Since $\ell n(2cz)$ is smaller than $2cz$ for large z, $\lim\limits_{y \to c} P(y^1,y^2,y^3)$ and $\lim\limits_{y \to c} P(y)$ are smaller than $\lim\limits_{z \to +\infty} (\text{constant})z^3 \exp\{-A[\ell n(2cz)]^2\}$. This limit equals zero due to Eqs.(4a-4b), so both $P(y^1,y^2,y^3)$ and $P(y)$ vanish at y=c. Seeing that

$$\frac{z^2 \exp\{-A[\ell n(2cz)]^2\}}{z^{-n}} \text{ or } \frac{z[\ell n(2cz)]^2 \exp\{-A[\ell n(2cz)]^2\}}{z^{-n}} < \frac{\exp\{-A[\ell n(2cz)]^2\}}{z^{-(n+3)}},$$

for large z and the limit on the right-hand side equals zero due to Eqs.(4a-4b), we know, as z goes to infinity, both $P(y^1,y^2,y^3)$ and $P(y)$ fall off to zero faster than $z^{-n}$. Again, seeing that

$$\frac{\exp\{-[2cz]^B\}}{z^2 \exp\{-A[\ell n(2cz)]^2\}} \text{ or } \frac{\exp\{-[2cz]^B\}}{z[\ell n(2cz)]^2 \exp\{-A[\ell n(2cz)]^2\}} < \frac{\exp\{-[2cz]^B\}}{\exp\{-A[\ell n(2cz)]^2\}}$$

for large z and the limit on its right-hand side equals zero due to Eqs.(5a-5b), we can assert that both $P(y^1,y^2,y^3)$ and $P(y)$ fall off to zero slower than $\exp\{-[2cz]^B\}$ as z goes to infinity.

The prediction on a new decay mode of high-energy tails in velocity and velocity rate distributions of astrophysical plasma particles is: These velocity and velocity rate distributions fall off to zero, as y goes to c, slower than any exponential decay, $\exp\{-[2c/(c-y)]^B\}$, but faster than any power-law decay, $(c-y)^n$, where B and n are two positive numbers.

ACKNOWLEDGMENT

The author greatly appreciates the teachings of Prof. Wo-Te Shen. The author thanks Prof. Gerhard Muller and Dr. P. Rucker for their supports of this work.

REFERENCES

[1]    E. T. Sarris et al, Geophys. Res. Lett., 8, 349 (1981)

       J. T. Gosling et al, J. Geophys. Res., 86, 547 (1981)

       W. K. Peterson et al, J. Geophys. Res., 86, 761 (1981)

       J. D. Scudder, Astrophys. J., 398, 99 (1992)

       V. Pierrard and J. Lemaire, J. Geophys. Res., 101, 7923 (1996)

       D. J. Williams et al, Geophys. Res. Lett., 15, 303 (1988)

       R. L. Mace, J. Geophys. Res., 103, 643 (1998)

       M. R. Collier, Geophys. Res. Lett., 20, 1531 (1993)

       D. T. Decker et al, J. Geophys. Res., 100, 21409 (1995)

       S. Xue, M. Thorne and D. Summers, J. Geophys. Res., 101, 15467 (1996)

       M. P. Leubner, J. Geophys. Res., 105, 21261 (2000)

[2]    S. P. Christon et al, J. Geophys. Res., 93, 2562 (1988)

[3]    V. M. Vasyliunas, J. Geophys. Res., 73, 2839 (1968)

       D. Summers and R. M. Thorne, Phys. Fluids, B3, 1835 (1991)

[4]    Jian-Miin Liu, Chaos, Solitons & Fractals, 12, 2149 (2001) [physics/0108045]; cond-mat/0108356; Relativistic corrections to the Maxwellian velocity distribution, to be published

[5]    Jian-Miin Liu, Chaos, Solitons & Fractals, 12, 399 (2001) [physics/0108044]; 12, 1111 (2001) [the revised version of hep-th/9805004]